\documentclass[aps,pra,twocolumn,groupedaddress,epsfig,amsmath,floatfix,showpacs]{revtex4}
\usepackage{epsfig,amsmath}
\usepackage{graphicx}
\usepackage{dcolumn}
\usepackage{bm}
\begin{document}


\title{Classical Effects of Laser Pulse Duration on Strong-field Double Ionization}
\author{Phay J. Ho and J. H. Eberly}
\affiliation{{Department of Physics and Astronomy,} \\
{University of Rochester, Rochester, NY 14627}}
\date{\today}

\begin{abstract}

We use classical electron ensembles and the aligned-electron approximation to examine the effect of laser pulse duration on the dynamics of strong-field double ionization. We cover the range of intensities $10^{14}-10^{16} W/cm^2$ for the laser wavelength 780 nm. The classical scenario suggests that the highest rate of recollision occurs early in the pulse and promotes double ionization production in few-cycle pulses.  In addition, the purely classical ensemble calculation predicts an exponentially decreasing recollision rate with each subsequent half cycle.  We confirm the exponential behavior by trajectory back-analysis.

\end{abstract}

\maketitle

Detailed study of the classical picture of two-electron and multi-electron ejection from atoms and molecules is becoming important as short-pulse and high-field experimental work moves into the multi-electron regime \cite{N-electronexpts}. This is because it is well recognized  that there is no realistic prospect of a quantum mechanical treatment of atomic and molecular multi-electron phenomena under the relevant experimental conditions, i.e., under highly non-perturbative fields with femtosecond and shorter time resolution. The exact solution of Schr\"odinger's time-dependent equation for helium double ionization in this regime by the Taylor group \cite{Taylor} is accepted as the farthest that a comprehensive quantum treatment will go in the foreseeable future. At the same time, evidence has been accumulating that a purely classical scenario \cite{Ho-etal05,Ho-PRA05} is consistent with many of the double ionization effects that are observed, and can be adapted further for multiple electrons.

Simultaneous ejection of two electrons from atoms more complex than helium has been under experimental study for more than two decades \cite{early experiments}, and has been studied intensely in the decade since the discovery and confirmation \cite{NSDI-atoms} of a knee-like structure in the experimental ion count measurements in several inert gases, now accepted as implying anomalously high electron correlation.  This signature has been reported for some molecules as well \cite{NSDI-molecules}.  Theoretical work based on a sequential ionization process involving independent electrons \cite{PPT-ADK} is known to be inadequate to produce the knee signature.

\begin{figure}
\centerline{\includegraphics[width=3.0in]{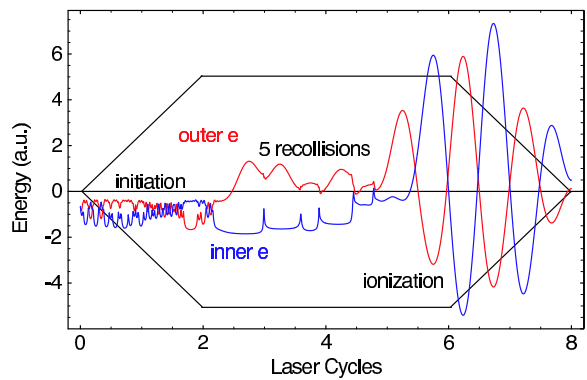}}
\caption{ (Color online)  Plots of electron energy versus time measured in laser cycles, showing a pair of two-electron trajectories, with the trapezoidal laser field envelope $f(t)$ superposed for reference. During laser turn-on the bound electrons exchange energy rapidly, and soon after laser turn-on is complete one (lighter red line) escapes and then returns in repeated e-e recollisions until the inner electron (darker blue line) also has enough energy to escape the nucleus, and thus the pair become doubly ionized.  Notice that after each recollision, the energy of the inner electron is modified, and the outer electron carries different energies in each recollision.}
\label{fig.energyDISI}
\end{figure}

The prevailing picture of double ionization builds on the familiar electron rescattering scenario proposed for NSDI by Corkum \cite{Corkum}.  In this picture, one electron is ionized by the Keldysh mechanism \cite{Keldysh-65}, tunneling out of the nucleus through a Coulomb barrier that has been lowered by the strong laser field.  Then the laser field returns it in a purely classical fashion to the nucleus to collide with the bound electron and free it. A relatively direct classical analysis, without tunneling, of the effect can be based on solutions of Newtonian equations for individual electron trajectories obtained from large ensembles of 100,000 - 1,000,000 members, which can be interpreted in a clear way \cite{Panfili-etal01}. An example of a pair of energy trajectories is shown in Fig. \ref{fig.energyDISI}, where a  sequence of rescatterings is clearly evident. Agreement between such classical and more elaborate partly or entirely quantum calculations \cite{smatrix, coulomb-focusing, softcollision, Texas} is remarkably good \cite{Panfili-etal02, Haan-etal02, Liu-Figueira04}.

Here we use the classical ensemble method \cite{Panfili-etal01} with 100,000 2-e trajectories and laser pulses with different numbers of laser cycles over a range of intensities, $10^{14}-10^{16}$ W/cm$^2$. Each laser pulse has the experimental wavelength 780 nm. Our laser field is taken to be a sine wave \cite{phasedependence} with maximum field ${\cal E}$ and a trapezoidal envelope function $1 \ge f(t) \ge 0$ with 2-cycle turn-on and turn-off, which is shown superposed on Fig. \ref{fig.energyDISI}.  For this report the number of cycles in the plateau ranges from 0 to 10.  The trapezoidal shape is chosen because it provides a fixed plateau intensity and is known to suppress effects associated solely with turn-on \cite{Grobe-Fedorov93}.

\begin{figure}
\centerline{\includegraphics[width=3.0in]{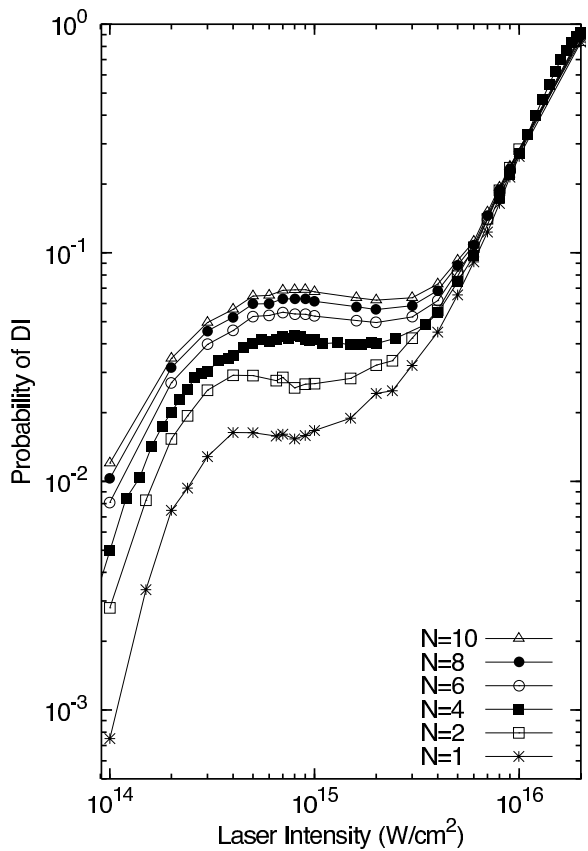}} \caption{Variation of ``knee" structure in laser pulses with different number of plateau cycles ($N$).  The probability of double ionization is the number of double ionization events divided by 100,000, the total number of 2-e trajectories used.} \label{fig.pulselength}
\end{figure}

To begin we show in Fig. \ref{fig.pulselength} that in the domain of interest here the knee signature of NSDI is preserved in a classical ensemble theory of the double ion count even for short pulses. We see that for intensities above the main knee region the double ionization yield is basically not dependent on the number of cycles in the pulse. We will see further evidence that in this regime the field is sufficiently strong to effect both electrons within a single cycle.

As Fig. \ref{fig.energyDISI} shows, while the pulse is turning on a rapidly repeated energy exchange occurs between the bound electrons, and this is what permits one to escape (without  tunneling) over the Coulomb barrier, which becomes more and more suppressed as the field turns on. Our ensemble calculations  \cite{Ho-etal05} establish that when many cycles of laser pulse are available Fig. \ref{fig.energyDISI} is typical of the classical process, i.e., the typical double-ionization event has been preceded by several or even many recollisions. This picture  stands in two-way contrast to the ``quantum" picture in which tunneling is employed to release the first electron, and where double ionization is deemed to occur in the very first recollision cycle.

There has been no way to choose between these two pictures of the double ionization timing. Experiments do not reveal how many cycles pass before the first electron is released. This uncertainty has begun to be eliminated by the ability to use shorter pulses \cite{shortpulseexpts}, which can suppress recollision effects requiring more than one cycle. These experiments show that double ionization does not require multiple recollisions,  consistent with the ``quantum" picture.  The open question is whether the classical theory, which has multiple-recollisions as typical, and doesn't invoke tunneling, agrees with this.  The preservation of the knee signature in short pulses as shown in Fig. \ref{fig.pulselength} confirms that NSDI can still proceed in the absense of the typical multiple-recollisions.  This positive answer prompts another question: how does this classical picture incorporate both of these long and short pulse duration NSDI schemes?   As we show here, the answer further confirms the attraction of a purely classical picture.

In order to accomplish classical double ionization the outer electron must find the inner electron in a suitable phase relation to its motion while going through the nuclear region. We have already demonstrated \cite{Panfili-etal02} the efficacy of well-phased (what were called ``slow-down") recollisions. Of course, the outer electron may or may not return to the nucleus to make a recollision.  If it returns to the nucleus, the recollision may take place in a condition that is more or less favorable than in the previous encounter. Because of the energy exchange that initially helped free the outer electron, the inner electron is generally to be found near the bottom of the nuclear potential well, where its bound oscillations are rapid and random. In a large ensemble it is reasonable that there will always be some in the right phase relation, even very early in the pulse.

\begin{figure*}
\centerline{\includegraphics[width=6.6in]{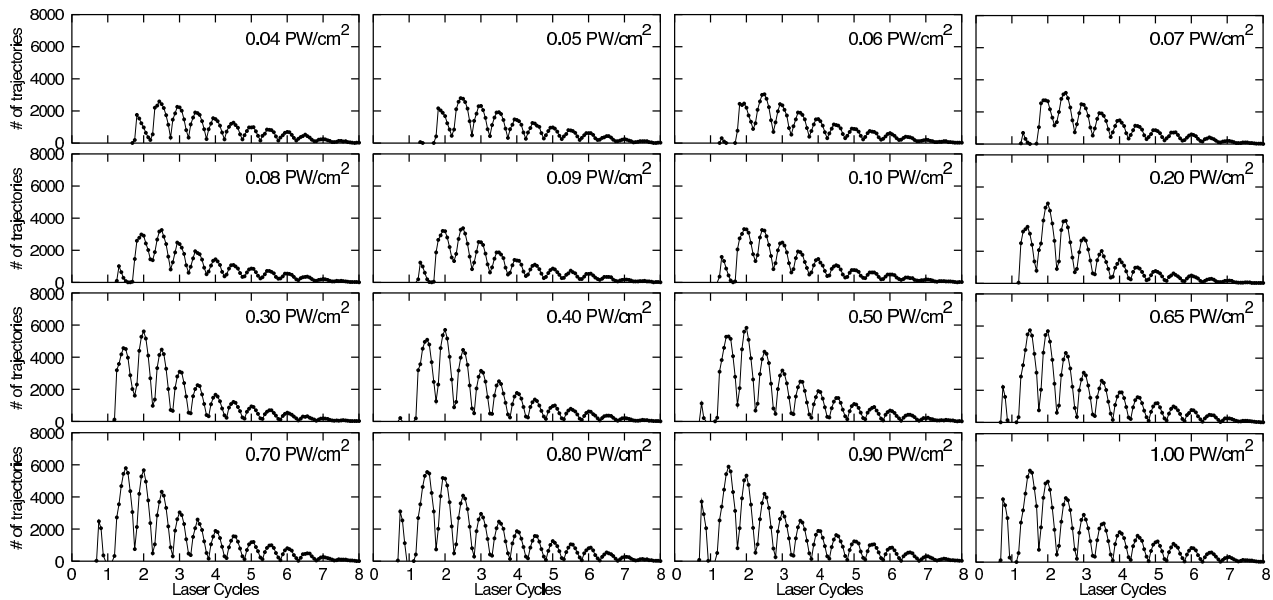}}
\caption{Number of recolliding electrons vs. time, counted at 8 times in each half cycle, for sixteen different intensities of 8-cycle pulses, from $4 \times 10^{13}$ to $10^{15}$ W/cm$^2$. The pulses with intensities in this range, which includes the main knee region, remove electron pairs smoothly and steadily, leading to the almost exactly exponential decrease of peak numbers through the duration of the pulse.}
\label{fig.returntimeA}
\end{figure*}

Our data confirms this qualitative picture, and provides features not available in a quantum calculation, i.e., data based on specific recollision counts. This can provide insights that are otherwise inaccessible. The first example is shown in Fig. \ref{fig.returntimeA}. Here we show a new form of time record for pulses with 16 different intensities. We count the number of trajectories that fit the ``recolliding" definition \cite{recollidingDef} at 256 time instants, i.e., 32 times per cycle, for 8-cycle pulses at 16 different intensities. In these records the intensities are low enough that the entire recollision sub-ensemble is smoothly depleted during the double ionization over 8 cycles, and one can anticipate a straightforward analysis, as follows.

\begin{figure}
\centerline{\includegraphics[width=3.4in]{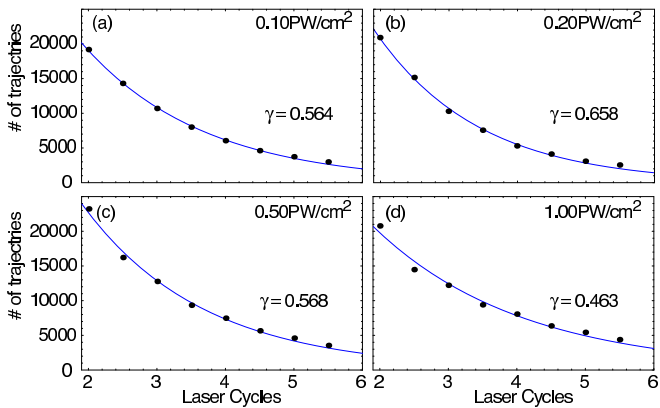}}
\caption{ (Color online)  (a)-(d) Total number of recolliding electron trajectories in the overall ensemble counted at eight half cycles of an 8-cycle trapezoidal pulse with a 2-cycle on-ramp and 2-cycle off-ramp. Data is shown for four laser intensities in the NSDI knee region, and intermediate intensities produce results in the same range.  The solid line in each plot corresponds to the best fit exponential function characterized with an exponential rate $\gamma$ for the data points. }
\label{fig.expondecay}
\end{figure}

With each successful recollision some of the available population is removed. This means that the next recollision has a fraction fewer candidates available for double ionization. A fractional decrease per half cycle is the same as predicting an exponential decrease in the number of recollisions over time. To make a check of this prediction the classical approach suggests use of our ``back-analysis" method \cite{Panfili-etal02}, but the number of successful double ionizations in any half cycle is very small even in a 100,000-member ensemble, making statistical reliability difficult to achieve. However, the number of recollisions, successful or not, is much greater and can be used as a surrogate for the number of successful ones. Fig. \ref{fig.expondecay} shows data extracted from Fig. \ref{fig.returntimeA} that confirms the exponential prediction and gives the exponential rates.

One clear feature quickly seen in this new form of data is that the half cycle that typically has the most recollisions, gateways to NSDI, is the very first one after the pulse reaches full strength.  By itself this establishes that the classical picture does not neccessarily require many cycles of recollision in order to generate NSDI electrons.  Since the more typical double ionization occurs after several or more cycles, the absence of these multiple-recollision gateways in short pulses reduces the production of double ionization as indicated in Fig. \ref{fig.pulselength}.  However, at only sightly higher intensities the picture changes.

\begin{figure*}
\centerline{\includegraphics[width=6.6in]{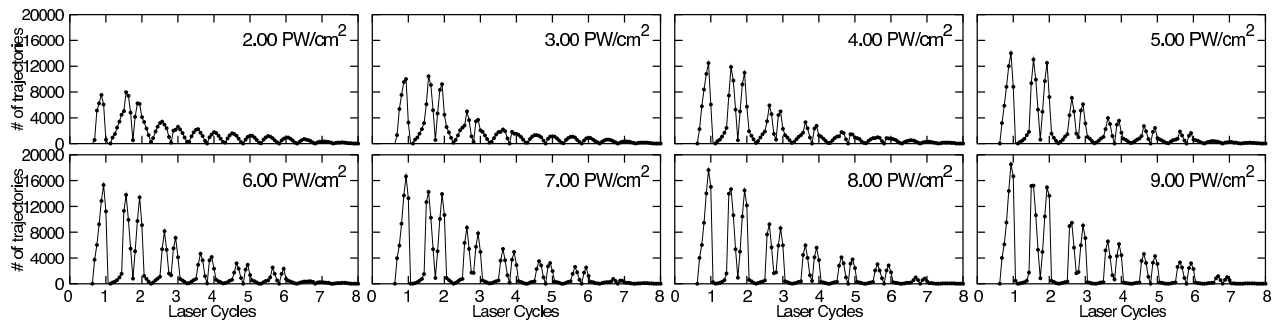}}
\caption{Number of recolliding electrons vs. time, counted at 8 times in each half cycle, for eight different intensities of 8-cycle pulses, from $2 \times 10^{15}$ to $9 \times 10^{15}$ W/cm$^2$. The intensities in this range begin to be high enough to distort the NSDI process. The doubled peaks indicate that a major fraction of all members of the ensemble can be induced to leave the nuclear region in the same direction at the same time.}
\label{fig.returntimeB}
\end{figure*}

In Fig. \ref{fig.returntimeB} the record shows the same type of data shown in Fig. \ref{fig.returntimeA}, but for a range of higher pulse intensities. The difference in effect is substantial, not only because the higher range of intensities promotes more effective double ionization, but because it also alters the nature of the process. The recollision begins to occur twice as rapidly. That is, it starts at the halfway point of the turn-on.  Furthermore, most recollisions occur prior to peak intensity.  The higher intensities in Fig. \ref{fig.returntimeB} merely emphasize this effect. One can see it beginning as early as in the third row of Fig. \ref{fig.returntimeA} where recollision starts prior to peak intensity.  At high intensities, this effect coupled with the finding that the double ionization yield does not depend much on the pulse duration suggests that NSDI events can be completed within a single cycle.  

In summary, we have shown that NSDI production is predicted by the purely classical theory to occur at short times well in advance of the typical double ionization time.   We see that the pulse begins to be strong enough to disturb a substantial fraction of the ensemble even before the plateau regime is reached at the beginning of the second cycle. In fact, this distortion is due to the stronger effect of barrier suppression, which promotes one of the bound electrons to leave the nucleus, and allows recollisions to take place early in the pulse, which substantially enhances the NSDI production from a short pulse.

This work was supported by DOE Grant DE-FG02-05ER15713.

\bibliography{apssamp}

\begin{thebibliography}{99}
\bibitem{recollidingDef}  In our calculation, a ``recolliding" trajectory at time $t$ satisfies two conditions: (1) it is a singly ionized trajectory at time $t-T/16$, where $T$ is the period of a laser cycle, and (2) the outer electron must cross the nucleus between $t-T/16$ and $t$.
\end{thebibliography}

\end{document}